\documentstyle[11pt,amssymb,epsf]{article}
\makeatletter
\@addtoreset{equation}{section}
\makeatother

\def\ben{\begin{equation}}
\def\een{\end{equation}}

  \let\n=\nu  \let\p=\pi

\let\C=\Chi

 \def\bd{\begin{document}} \def\ed{\end{document}}
\def\ds{\documentstyle} \let\fr=\frac \let\bl=\bigl \let\br=\bigr
\let\Br=\Bigr \let\Bl=\Bigl
\let\bm=\bibitem
\let\na=\nabla
\let\pa=\partial \let\ov=\overline
\newcommand{\be}{\begin{equation}}
\newcommand{\ee}{\end{equation}}
\def\ba{\begin{array}}
\def\ea{\end{array}}
\def\ft#1#2{{\textstyle{\frac{\scriptstyle #1}{\scriptstyle #2} } }}
\def\fft#1#2{{\frac{#1}{#2}}}
\def\del{\partial}
\def\vp{\varphi}
\def\sst#1{{\scriptscriptstyle #1}}
\def\oneone{\rlap 1\mkern4mu{\rm l}}
\def\td{\tilde}
\def\wtd{\widetilde}
\def\ie{{\it i.e.\ }}
\def\dalemb#1#2{{\vbox{\hrule height .#2pt
        \hbox{\vrule width.#2pt height#1pt \kern#1pt
                \vrule width.#2pt}
        \hrule height.#2pt}}}
\def\square{\mathord{\dalemb{6.8}{7}\hbox{\hskip1pt}}}
\newcommand{\ho}[1]{$\, ^{#1}$}
\newcommand{\hoch}[1]{$\, ^{#1}$}
\newcommand{\bea}{\begin{eqnarray}}
\newcommand{\eea}{\end{eqnarray}}
\newcommand{\ra}{\rightarrow}
\newcommand{\lra}{\longrightarrow}
\newcommand{\Lra}{\Leftrightarrow}
\newcommand{\bp}{\tilde \beta^\prime}
\newcommand{\tr}{{\rm tr} }
\newcommand{\Tr}{{\rm Tr} }
\def\0{{\sst{(0)}}}
\def\1{{\sst{(1)}}}
\def\2{{\sst{(2)}}}
\def\3{{\sst{(3)}}}
\def\4{{\sst{(4)}}}
\def\5{{\sst{(5)}}}
\def\6{{\sst{(6)}}}
\def\7{{\sst{(7)}}}
\def\8{{\sst{(8)}}}
\def\n{{\sst{(n)}}}
\def\cA{{{\cal A}}}
\def\cB{{{\cal B}}}
\def\cF{{{\cal F}}}
\def\cH{{{\cal H}}}
\def\tV{\widetilde V}
\def\tW{\widetilde W}
\def\tH{\widetilde H}
\def\tE{\widetilde E}
\def\tF{\widetilde F}
\def\tA{\widetilde A}
\def\im{{{\rm i}}}
\def\tY{{{\wtd Y}}}
\def\ep{{\epsilon}}
\def\vep{{\varepsilon}}
\def\bD{{{\bar D}}}
\def\R{{{\mathbb R}}}
\def\C{{{\mathbb C}}}
\def\H{{{\mathbb H}}}
\def\CP{{{\mathbb C}{\mathbb P}}}
\def\RP{{{\mathbb R}{\mathbb P}}}
\def\Z{{{\mathbb Z}}}
\def\bA{{{\mathbb A}}}
\def\bB{{{\mathbb B}}}
\def\bC{{{\mathbb C}}}
\def\bD{{{\mathbb D}}}
\def\bE{{{\mathbb E}}}
\def\bZ{{{\mathbb Z}}}
\def\Re{{{\frak{Re}}}}
\def\Im{{{\frak{Im}}}}
\def\cosec{{\,\hbox{cosec}\,}}
\def\Gm{{\Gamma_{\!\! -}}}
\def\Gp{{\Gamma_{\!\! +}}}
\def\stan{{standard }}
\def\nonstan{{supernumerary }}
\def\p{{\partial}}
\def\kdel#1{{\fft{\del}{\del#1}}}

\def\bog{{Bogomolny }}

\def\beps{{\bar\epsilon}}
\newcommand{\ww}[1]{\\[0.#1cm]}
\def\eps{\epsilon}
\def\slashchar#1{\setbox0=\hbox{$#1$}           % set a box for #1
   \dimen0=\wd0                                 % and get its size
   \setbox1=\hbox{/} \dimen1=\wd1               % get size of /
   \ifdim\dimen0>\dimen1                        % #1 is bigger
      \rlap{\hbox to \dimen0{\hfil/\hfil}}      % so center / in box
      #1                                        % and print #1
   \else                                        % / is bigger
      \rlap{\hbox to \dimen1{\hfil$#1$\hfil}}   % so center #1
      /                                         % and print /
   \fi}
\def\sd{\slashchar{D}}
\def\sp{\slashchar{\partial}}
\newcommand{\eq}[1]{(\ref{#1})}
\def\cdh{\Gamma\cdot H}
\def\cL{e^{-1} {\cal L}}

\begin{document}

\begin{flushright}
\hfill{ \
CAS-KITPC/ITP-204\ \ \ \ }
% \hfill{
%\bf hep-th/yymmnnn}
\end{flushright}

\vspace{25pt}
\begin{center}
{\large {\bf Thermodynamical Metrics and Black Hole Phase Transitions}}

\vspace{15pt}

Haishan Liu\hoch{1,2}, H. L\"u\hoch{3,4}, Mingxing Luo\hoch{1} and
Kai-Nan Shao\hoch{1,2}

\vspace{10pt}

\hoch{1}{\it Zhejiang Institute of Modern Physics, Department of
Physics\\
Zhejiang University, Hangzhou, 310027}

\vspace{10pt}

\hoch{2}{\it Kavli Institute for Theoretical Physics China, CAS, Beijing 100190, China}

\vspace{10pt}

\hoch{3}{\it China Economics and Management Academy\\
Central University of Finance and Economics, Beijing 100081}

\vspace{10pt}

\hoch{4}{\it Institute for Advanced Study, Shenzhen
University,
Nanhai Ave 3688, Shenzhen 518060}

\vspace{40pt}

\underline{ABSTRACT}
\end{center}

An important phase transition in black hole thermodynamics is
associated with the divergence of the specific heat with fixed
charge and angular momenta, yet one can demonstrate that neither
Ruppeiner's entropy metric nor Weinhold's energy metric reveals this
phase transition. In this paper, we introduce a new thermodynamical
metric based on the Hessian matrix of several free energy. We
demonstrate, by studying various charged and rotating black holes,
that the divergence of the specific heat corresponds to the
curvature singularity of this new metric. We further investigate
metrics on all thermodynamical potentials generated by Legendre
transformations and study correspondences between curvature
singularities and phase transition signals. We show in general that
for a system with $n$-pairs of intensive/extensive variables, all
thermodynamical potential metrics can be embedded into a flat
$(n,n)$-dimensional space. We also generalize the Ruppeiner metrics
and they are all conformal to the metrics constructed from the
relevant thermodynamical potentials.

\vspace{15pt}

\thispagestyle{empty}

%\pagebreak
%\voffset=0pt
%\setcounter{page}{1}

%\tableofcontents

%\addtocontents{toc}{\protect\setcounter{tocdepth}{2}}

\newpage

%%%%%%%%%%%%%%%%%%%%%%%%%%%%%%%%%%%%%%%%%%%%%%%%%%%%%%%%%%%%%%%%%%%

\section{Introduction}

%%%%%%%%%%%%%%%%%%%%%%%%%%%%%%%%%%%%%%%%%%%%%%%%%%%%%%%%%%%%%%%%%%%

Recently there has been considerable interest in applying the
AdS/CFT correspondence \cite{adsref} to understand certain aspects
of condensed matter physics. In these studies, non-extremal black
holes that are asymptotically anti-de Sitter (AdS) are the most
natural backgrounds in the gravitational dual. This is due to the
fact that a condensed matter system usually has a non-vanishing
temperature, while gravitational configurations with temperature and
appropriate asymptotic behaviors always lead to non-extremal black
holes.

For asymptotically Minkowski spacetime, there can be one and only
one Schwarzschild black hole at a given temperature. It has negative
specific heat and evaporates via Hawking radiation. For an AdS
Schwarzschild black hole, the story is different. The boundary of
the AdS spacetime acts like a thermal box \cite{york}, and as such,
black holes can only emerge above certain minimum temperatures.
Above these minimum temperatures, there can be two types of black
holes: the small ones are much like the usual Schwarzschild black
hole of negative specific heat, whilst the big ones have positive
specific heat and are locally stable. With the increasing of the
temperature, the Helmoltz free energy of the large black holes can
become less than the thermal background. This implies that above
certain temperature pure thermal radiation in AdS becomes unstable
and collapses to form black holes. This phase transition from the
AdS thermal radiation to a black hole phase was discovered by
Hawking and Page \cite{hawkingpage}. It turns out that the
Hawking-Page phase transition can be interpreted, {\it via} the
AdS/CFT correspondence, as a transition from a low-temperature
confining phase to a high temperature de-confining phase in the
boundary field theory \cite{witphase}.

It is natural to generalize the AdS Schwarzschild black hole to
higher dimensions and to include charges and angular momenta. The
thermodynamical properties of the Reinssner-Nordstr\"om (RN) AdS
black holes in arbitrary dimensions were studied in \cite{cejm}, and
those of the Kerr-Newman-AdS black holes in \cite{cck}. It turns out
that in those cases, locally stable black holes with positive
specific heat exist for all temperature, including zero temperature
which corresponds to the extremal limit. Increasing the temperature
from zero, the (locally-stable) small black hole can undergo a
first-order phase transition to become a large black hole, for
charges or angular momenta less than certain critical values. This
thermodynamical phase transition is different from the global
Hawking-Page phase transition and it is characterized by the
divergence of thermodynamical linear response function such as the
specific heat. In this paper, we shall consider mainly this type of
local thermodynamical phase transitions.

When the number of conserved quantities such as charges and angular
momenta increases, the phase transitions become more and more
complicated.  In gauged supergravities, large classes of
non-extremal charged rotating AdS black holes have been constructed
\cite{clp,cclp,cclp1,chow}. It would then be extremely useful to have a
relative simple mechanism to identify phase transition points.
One way is to study the thermodynamical geometry on the state space.

An important question now is how to construct metrics on this space.
The simplest construction is based on the Hessian matrix of a
certain quantity such as a thermodynamical potential, which is a
function of either temperature or entropy together with other
extensive variables. Consider a function $f$ with variables $x^i$,
$i=1,\cdots, n$, the elements of the corresponding Hessian matrix
are $h_{i,j}=\partial^2f/(\partial x^i\partial x^j)$.  One can then
construct a geometry with $x^i$ as coordinates and $h_{i,j}$ as the
metric components, {\it i.e.}
\begin{equation}
ds^2=g_{ij} dx^i dx^j\,,\qquad \hbox{with}\qquad g_{ij}=h_{i,j}\,.
\end{equation}
The first of such a metric in thermodynamics was introduced by
Weinhold \cite{wein}, making use of the first law of thermodynamics
$dM=T dS + \sum \mu_i dN_i$, as the energy $M$ is naturally a
function of the entropy and extensive variables $N_i$. Ruppeiner
later introduced a different metric, treating the entropy as the
generating function depending on $M$ and $N^i$ \cite{rupp}. At the
first sight, Ruppeiner's entropy metric would appear to be most
appropriate for black holes since the entropy, evaluated on the
horizon, is rather different from the energy and $N_i$, which are
all evaluated at the asymptotic infinity. However, it turns out that
the Ruppeiner metric is related to the Weinhold's energy metric by
an overall conformal factor $1/T$ \cite{ruppwein}. There have been
considerable amount of works in understanding these thermodynamical
geometries of black holes, in particular to space-times which
asymptote to AdS \cite{fgk}-\cite{latest}.

It should be mentioned right away that any geometry can only be
non-trivial with at least two dimensions. However, the Hawking-Page
phase transition of the AdS Schwarzschild black hole occurs in a
one-dimensional system with only energy as its solution variable.
This serves as a warning about the limitation of thermodynamical
metrics. Nevertheless, studies of black holes with additional
extensive variables have all painted a tantalizing picture relating
curvature singularities of Ruppeiner or Weinhold metrics to phase
transition points.

The signal of a phase transition typically appears when a capacity,
such as the specific heat, charge capacitance or moment of inertia,
changes the sign, which implies a change of stability. The sign
change of these quantities can only happen by going through either
zero or infinity. It was shown in \cite{rupp2} that whilst the
Ruppeiner and Weinhold metrics indeed reveal the signals of black
hole phase transitions associated with divergence of specific heat
with fixed electric potential or angular velocity, they are
insensitive to the Davies curve \cite{daviescurve} where the
specific heat with fixed charge and/or angular momentum diverges.
Various {\it ad hoc} modifications for the geometries were proposed
\cite{Shen:2005nu,Alvarez:2008wa}.

     In this paper, we introduce new metrics based on the Hessian
matrix of all thermodynamical potentials generated by Legendre
transformations of the black hole energy or entropy. We calculate
the corresponding Ricci scalars for various black holes and find
that the collection of all curvature singularities are in one-to-one
correspondence to the collection of threshold points of all
capacities. In particular we find that the free-energy metric is
always associated with the Davies curve.

The paper is organized as follows.  In section 2, we present a
general discussion of thermodynamical metrics.  We show that all
thermodynamical potential metrics including the Weinhold metric can
be embedded into a higher-dimensional flat space. We also generalize
the Ruppeiner metric and all these generalizations are conformal to
the relevant potential metrics. In section 3, we consider
Reissner-Nordstr\"om black holes in general dimensions. By examining
all capacities including both specific heat $C_Q$ and $C_\Phi$ and
both charge capacitances $\widehat C_T$ and $\widehat C_S$, we find
that there are two threshold points where these quantities change
sign by going through either infinity or zero. We calculate the
Ricci scalar curvature for both Weinhold and Ruppeiner metrics and
they share one same singularity that is precisely located at one of
the threshold points. However, these two metrics are insensitive to
the second threshold point. This prompts us to derive the Ricci
scalar of the free-energy metric, and we find that its singularity
corresponds to precisely the second threshold point. It is natural
then to consider metrics generated by all thermodynamical potentials
obtained by Legendre transformations. We find that they form
conjugate pairs, in which the metrics are negative of each other.
For RN AdS black holes, curvature singularities of Weinhold and
free-energy metrics, obtained {\it via} Legendre transformations of
the energy, give rise the to full set of threshold points.

In section 4, we consider the Kerr-Newman-AdS black hole in four
dimensions.  In this case, there are a total of twelve capacity-type
quantities and eight thermodynamical potentials. We find that the
full collection of threshold points of twelve capacities is exactly
the same as that of curvature singularities of metrics
associated with the eight potentials.

In section 5, we consider Kerr-AdS black holes.  Owing to the
complexity in the general case, we consider only four special
examples. These are $D=4$ and $D=5$ solutions and general
dimensional solutions of specific angular momenta. In section 6, we
consider the Ricci-flat black ring solution with single angular
momentum. Although details may be different in each case, essential
features are of the same. We conclude our paper in section 7.  In
appendix A, we examine the van der Waals model and demonstrate that,
as in the cases of black hole thermodynamics, additional metrics are
needed and all threshold points can indeed be revealed. In
appendices B and C, we present detailed results for discussions in
sections 4 and 5, respectively.

%%%%%%%%%%%%%%%%%%%%%%%%%%%%%%%%%%%%%%%%%%%%%%%%%%%%%%%%%%%%%%%%%%%

\section{General properties of thermodynamical geometry}

%%%%%%%%%%%%%%%%%%%%%%%%%%%%%%%%%%%%%%%%%%%%%%%%%%%%%%%%%%%%%%%%%%%

In thermodynamical geometries, it is important to construct an
appropriate metric for the equilibrium state space of a
thermodynamical system.  Let us start with the first law of
thermodynamics
\begin{equation}
dM = T dS + \sum_{i=1}^n \mu_i dN_i\,.
\end{equation}
Note that here we have adopted the black hole notations and used $M$
to denote the ``internal'' energy of the system.  The system has
$(n+1)$ pairs of intensive/extensive variables $(T,S)$ and $(\mu_i,
N_i)$.  Here $(\mu_i, N_i)$ can be pairs of pressure/volume $(P,V)$,
electric potential/charge $(\Phi,Q)$, or angular velocity/momemtum
$(\Omega, J)$, {\it etc}. The whole thermodynamical state space can
be viewed as a $2(n+1)$-dimensional embedding of the
$(n+1)$-dimensional space.  The energy $M$ is a function of $(n+1)$
extensive variables $(S, N_i)$.  We find that the energy metric
based on the Hessian matrix introduced by Weinhold can be simply
written as
\begin{equation}
ds^2(M) = dT dS + \sum_{i=1}^n d\mu_i dN_i\,.\label{weinmet}
\end{equation}
For black hole theromodynamics, this is a particularly convenient
way to express the metric. This is because although the natural
variables for $M$ are $S$ and $N_i$, the thermodynamical quantities
in black holes are typically expressed in terms of parametric
variables including the horizon radius. The explicit function
of $M$ in terms of $S$ and $N_i$ is not always available for a
generic black hole. A metric like (\ref{weinmet}) removes the
necessity of writing such an explicit function.

    The first law of thermodynamics can also be expressed as
\begin{equation}
dS = \fft{1}{T} dM - \sum_{i=1}^n \fft{\mu_i}{T} dN_i\,.
\end{equation}
It follows that the Ruppeiner entropy metric is given by
\begin{eqnarray}
ds^2(S) &=& d(\fft{1}{T}) dM - \sum_{i=1}^n d(\fft{\mu_i}{T}) dN_i
\cr %%
&=& - \fft{1}{T} ds^2(M)\,.
\end{eqnarray}
as was demonstrated in \cite{ruppwein}.

We now generalize the Weinhold energy metric to those of other
thermodynamical potentials yielded by Legendre transformations of
internal energy $M$. For example, we may consider the Helmholtz free
energy $F=M - TS$, as a function of $T$ and $N_i$, which satisfies
\begin{equation}
dF=-S dT + \sum_{i=1}^n \mu_i dN_i\,.
\end{equation}
The corresponding metric is given by
\begin{equation}
ds^2(F) = - dT dS + \sum_{i=1}^n d\mu_i dN_i\,.
\end{equation}
Thus we can use parametric variables for thermodynamical
quantities to write both the energy and free-energy metrics. The
difference of these two metrics is simply a sign change of the term
$dT dS$. This clearly simplifies the construction of metrics, by
bypassing the need of obtaining explicit functions $M(S,N_i)$ and
$F(T,N_i)$. By considering all possible thermodynamical potentials,
we arrive at a total of $2^{n+1}$ metrics, given by
\begin{equation}
ds^2= \pm dT dS + \sum_{i=1}^n \pm d\mu_i dN_i\,.\label{allmetrics}
\end{equation}
Here the plus and minus signs are independent. Since the overall
``$-$'' factor is trivial, there are $2^n$ inequivalent metrics.  We
call thermodynamical potentials $(U, \bar U)$ a conjugate pair if
they satisfy
\begin{equation}
U + \bar U = 2M - TS - \sum_{i=1}^n \mu_i N_i\,.
\end{equation}
Their associated metrics are negative of each other, {\it i.e.}
\begin{equation}
ds^2(U) = - ds^2 (\bar U)\,.
\end{equation}

It is worth emphasizing that (\ref{allmetrics}) implies that all
thermodynamical potential metrics can be embedded in the
$(n+1,n+1)$-dimensional space, with light-cone coordinates $(T,S)$
and $(\mu_i, N_i)$. The thermodynamical geometries are
$(n+1)$-dimensional hyper-surfaces in this $(n+1,n+1)$-dimensional
flat space. The metrics (\ref{allmetrics}) represent the totality of
all possible flat embeddings. The sign choices in
(\ref{allmetrics}), except for the overall one, are non-trivial and
cannot be absorbed by redefining the variables since it would imply
different embedding functions.

Generalizations of the Ruppeiner metric are straightforward. They
are all conformal to one of the potential metrics.  For example, let
us consider a thermodynamical potential $U$, satisfying
\begin{equation}
dU=\sum_{\alpha=0}^n \tilde \mu_\alpha d\tilde N_\alpha\,.
\end{equation}
The $U$ metric is given by $ds^2(U) = \sum_{\alpha} d\tilde
\mu_\alpha d\tilde N_\alpha$.  The Ruppeiner-like metric based on
the Hessian matrix on the function $\tilde N_\beta$ is then given by
\begin{equation}
ds^2(\tilde N_\beta) = - \fft{1}{\tilde\mu_\beta} ds^2(U)\,.
\end{equation}
Note that the natural variables for $\tilde N_\beta$ are $U$ and
$N_\alpha$'s with $\alpha\ne\beta$.

Having presented the general discussion of the thermodynamical
metrics, we shall study some specific examples in the following
sections.

%%%%%%%%%%%%%%%%%%%%%%%%%%%%%%%%%%%%%%%%%%%%%%%%%%%%%%%%%%%%%%%%%%%

\section{Reissner-Nordstr\"om AdS black holes}

%%%%%%%%%%%%%%%%%%%%%%%%%%%%%%%%%%%%%%%%%%%%%%%%%%%%%%%%%%%%%%%%%%%

Let us consider the Einstein-Maxwell theory with a negative cosmological
constant in $d$ dimensions.  The Lagrangian is given by
\begin{equation}
{\cal L}=\sqrt{-g}\Big(R - \ft14 F^2 + (d-1)(d-2)\lambda^2\Big)\,,
\end{equation}
where $F=dA$ is the field strength for the $U(1)$ vector potential
$A$. Here, we decided to parameterized the cosmological constant by
the inverse AdS radius, $\lambda \equiv \fft{1}{\ell}$. The theory
admits an electrically-charged solution, often denominated in the
literature Reissner-Nordstr\"om AdS black hole, given by
\begin{eqnarray}
ds^2&=&-V\, dt^2 + \fft{dr^2}{V} + r^2 d\Omega_{d-2}^2\,,\qquad A=
\fft{q\nu}{r^{d-3}} dt\,,\cr %%
V&=&1+\lambda^2r^2 - \fft{m}{r^{d-3}} +
\fft{q^2}{r^{2(d-3)}}\,,\qquad \nu=\sqrt{\fft{2(d-2)}{d-3}}\,.
\end{eqnarray}
The mass $m$ and the charge $q$ are conserved quantities. The
horizon of the black hole is located at $r=r_0$, which is the
largest real root of $V$. For later convenience, we may express $m$
in terms of $r_0$,
\begin{equation}
m=\fft{q^2}{r_0^{d-3}} + r_0^{d-3} (1 + \lambda^2 r_0^2)\,.
\end{equation}
It is straightforward to obtain thermodynamical quantities
such as the Hawking temperature, entropy, electric potential,
charge and mass. They are given by
\begin{eqnarray}
T&=&\fft{1}{4\pi}\Big(\fft{d-3}{r_0} - \fft{(d-3)q^2}{r_0^{2d-5}} +
(d-1)\lambda^2 r_0\Big)\,,\qquad S=\ft14 r_0^{d-2}
\omega_{d-2}\,,\cr%%
\Phi&=&\fft{q\nu}{r_0^{d-3}}\,,\qquad
Q=\fft{(d-3)q\nu\omega_{d-2}}{16\pi}\,,\cr%%
M&=&\fft{(d-2)(q^2r_0^6 + r_0^{2d}(1 + \lambda^2 r_0^2))
\omega_{d-2}}{16\pi r_0^{d+3}}\,.\label{rngenthermo}
\end{eqnarray}
Here $\omega_{d-2}=2 \pi^{(d-1)/2}/(\fft12 (d-3))!$ is a pure
numerical factor measuring the volume of the unit round $(d-2)$-sphere.
Note that these quantities satisfy the first law of thermodynamics,
\begin{equation}
dM=TdS + \Phi dQ\,.
\end{equation}

    An important quantity in thermodynamics is the specific heat defined
by $C\equiv T(\partial S/\partial T)$.  For
Tangherlini-Schwarzschild black holes, corresponding to $q=0$ and
$\lambda=0$ in our case, the specific heat is given by
$C=-\ft14(d-2) r_0^{d-2}\omega_{d-2}$, which is always negative and
signaling a local thermodynamical instability. For AdS Schwarzschild
black holes, it is given by
\begin{equation}
C=\fft{(d-2)\pi r_0^{d-1}\omega_{d-2} T}{(d-1)\lambda^2 r_0^2 -
(d-3)}\,.\label{hpheat}
\end{equation}
There is a minimal temperature $T_{\rm min}=\ft1{2\pi} \lambda
\sqrt{(d-1)(d-3)}$, corresponding to $\lambda^2 r_0^2 =
(d-3)/(d-1)$, above which black holes can tunnel to existence. This
is also a temperature at which the above specific heat diverges.
For a given temperature higher than $T_{\rm min}$, there could be
two types of black hole solutions, the small ones with $\lambda^2
r_0^2 < (d-3)/(d-1)$ have negative specific heat, whilst the big
ones with $\lambda^2 r_0^2 > (d-3)/(d-1)$ have positive specific
heat and are locally stable.  The Helmoltz free energy is
given by
\begin{equation}
F=\fft{1}{16\pi} r_0^{d-3} (1-\lambda^2 r_0^2)\,.
\end{equation}
When the temperature is above $T=\ft1{2\pi} (d-2)\lambda$,
corresponding to $\lambda r_0 > 1$, the large locally-stable black
holes start to have negative free energy and hence are more probable
than the pure AdS radiation \cite{hawkingpage}.

     For general RN AdS black holes, their phase transitions
were studied in detail in \cite{cejm}.  Owing to the existence of
charges, there can be extremal and near-extremal black holes, which
are locally stable. Locally stable RN black holes can exist at all
temperature.  When the temperature increases, small black holes can
have a phase transition to become a large black hole with
discontinuous entropy.  This is rather different from the global
Hawking-Page phase transition and is characterized by the divergence
of the specific heat.  To be specific, one may define the specific
heat for either constant charge $Q$ or constant electric potential
$\Phi$, given by
\begin{eqnarray}
C_Q &\equiv& T \fft{\partial S}{\partial T}\Big |_{Q} =
\fft{\partial M}{\partial T}\Big|_{Q} = \fft{(d-2)\pi r_0^{d-1}
\omega_{d-2} T}{\zeta_1}\,.\label{rncq}\cr %%
C_\Phi &\equiv & T\fft{\partial S}{\partial T}\Big |_\Phi = -
\fft{(d-2)\pi r_0^{3d-1}\omega_{d-2} T }{\zeta_2}\,.\label{rncphi}
\end{eqnarray}
where
\begin{eqnarray}
\zeta_1 &=& (d-1)\lambda^2 r_0^2 - (d-3)(1 - (2d-5) r_0^{2(3-d)}
q^2) \,,\cr %%
\zeta_2 &=& r_0^{2d} (d-3 - (d-1) \lambda^2 r_0^2) - (d-3) q^2 r_0^6
\,.
\end{eqnarray}
One may also introduce analogously charge capacitances
at fixed temperature or entropy, given by
\begin{eqnarray}
\widetilde C_T &\equiv& \fft{\partial Q}{\partial \Phi}\Big |_T =
\fft{(d-3)r_0^{3(d-1)} \omega_{d-2} \zeta_1}{16\pi \zeta_2}\,,\cr %%
\widetilde C_S &\equiv& \fft{\partial Q}{\partial \Phi}\Big |_S =
\ft{1}{16\pi} (d-3) r_0^{d-3} \omega_{d-2}\,.\label{rnc3}
\end{eqnarray}
We see that $\widetilde C_S$ is positive definite while the other three
$C$'s may change signs. There are two threshold points,
\begin{equation}
\zeta_1=0\,,\qquad {\rm and} \qquad \zeta_2=0\,.
\end{equation}
For non-vanishing $q$, $\zeta_1$ and $\zeta_2$ cannot vanish
simultaneously.  For zero cosmological constant, the condition for
$\zeta_2=0$ coincides with $T=0$.  The solution becomes extremal and
the mass/charge relation is saturated, {\it i.e.}~$M=M_{\rm
min}\equiv\nu Q$.  The condition for $\zeta_1=0$ implies that
\begin{equation}
M=\sqrt{\fft{2(d-2)^3}{(d-3)(2d-5)}}\, Q>M_{\rm min}\,.
\end{equation}

These threshold points are of two types.  For one type, the
corresponding specific heat and/or charge capacitance changes sign
by going through infinity. This is the case for $C_\Phi$ and
$\widetilde C_T$ at $\zeta_2=0$, and also $C_Q$ at $\zeta_1=0$. For
the other type, a capacity changes sign by going through zero, as
$\widetilde C_T$ at $\zeta_1=0$. For all black hole examples we have
examined in this paper, the vanishing of a certain capacity is
always associated with the divergence of another. Thus we shall not
be very strict on distinguishing the divergent or vanishing points.
As we shall see in section 6, the situation for the five-dimensional
black ring is somewhat different.

   Such threshold points signal the change of stability and hence
are typically associated with certain phase transitions. The phase
transition associated with the divergence of $C_Q$ was well studied
in \cite{cejm}.  A new type of phase transition for the RN black
hole associated with the divergence of $C_\Phi$ was recently
suggested in \cite{latest2,latest}.  It is natural to expect that
these important phase-transition points can be seen in a
thermodynamical metric, as curvature singularities.  As discussed in
section 2, the Weinhold and Ruppeiner metrics can be written as
\begin{equation}
ds^2(M)=dT dS + d\Phi dQ\,,\qquad
ds^2(S) = -\fft{1}{T} ds^2 (M)\,.
\end{equation}
Their Ricci scalars can be readily calculated, given by
\begin{eqnarray}
R^{(M)} &=&  \fft{16\pi (d-3)^2 r_0^{3(d+1)}}{(d-2) \omega_{d-2}
\zeta_2^2}\,,\cr %%
R^{(S)} &=& \fft{(d-1)\lambda^2 R^{(M)}}{16\pi^2(d-3)^2 r^{4d}
T}\Big( 3(d-3) q^2 r_0^6 - r_0^{2d}(3(d-3) - (d-1)\lambda^2
r_0^2)\Big)\times\cr &&\Big((d-3)(d-2)q^2 r_0^6 +
r_0^{2d}((d-3)(d-4) + (d-1)(d-2)\lambda^2 r_0^2)\Big)\,.
\end{eqnarray}
We see that the curvature singularity of these metrics are related
to the phase transition associated with the vanishing of $\zeta_2$,
but not $\zeta_1$. In Ruppeiner geometry, there is an additional
curvature singularity at $T=0$, corresponding to the extremal limit
of the black holes.  It is worth remarking that when the
cosmological constant parameter $\lambda=0$, the Ruppeiner curvature
vanishes and hence reveals no information at all.

 The inability of the Weinhold and Ruppeiner metrics to probe
the phase transition associated with the vanishing of $\zeta_1$
prompts us to look for other types of thermodynamical geometries. We
construct a new metric based on the Hessian matrix of the free
energy, defined by
\begin{equation}
F(T,Q)=M - T S\,,\qquad dF=-S dT + \Phi dQ\,.
\end{equation}
For RN AdS black holes, we have
\begin{equation}
F=\fft{(2d-5) q^2 r_0^6 + r_0^{2d} (1-\lambda^2 r_0^2)}{16\pi
r_0^{d+3}} \,.
\end{equation}
As discussed in section 2, it is not always convenient nor necessary
to use natural variables $(T,Q)$ to construct the free-energy
metric. We shall use the $(r_0,q)$ variables and the
flat-space embedding metric,
\begin{equation}
ds^2(F) = - dT\, dS + d\Phi\, dQ\,.
\end{equation}
Substituting the thermodynamical quantities in (\ref{rngenthermo}),
we obtain the free-energy metric with $(r_0,q)$ coordinates. The
Ricci scalar for the free-energy metric is then given by
\begin{equation}
R^{(F)}= \fft{16(d-3)^2 \pi r_0^{3(d+1)} (d-2 - (d-1)\lambda^2
r_0^2)}{(d-2) \omega_{d-2} \zeta_1^2}\,.
\end{equation}
The curvature singularity corresponds precisely to the phase
transition at $\zeta_1=0$, associated with the divergence of the
specific heat $C_Q$.

   We have also investigated other Legendre transformations, namely
$\bar F(S,\Phi)=M- \Phi Q$ and $\bar M(T,\Phi)=M - TS - \Phi Q$. As
discussed in section 2, they are conjugates to $F$ and $M$
respectively. It follows then
\begin{equation}
R^{(\bar F)} = - R^{(F)}\,,\qquad R^{(\bar M)} = - R^{(M)}\,.
\end{equation}
There are no further curvature singularities by considering these
additional metrics. This is encouraging since we do not expect any
further phase transitions in the RN AdS black holes. The totality of
curvature singularities of all metrics is exactly the same as the
totality of capacity threshold points. In \cite{Shen:2005nu}, a
generalized Ruppeiner metric was considered with variables $(M,Q)$
replaced by $(M-\Phi Q, \Phi)$. It follows from the discussion in
section 2, that metric is conformal to our $ds^2(F)=-ds^2(\bar F)$
with a conformal factor $1/T$. It has the same singularity as the
one associated with the divergence of $C_Q$.

It should be remarked that although thermodynamical geometries
reveal precisely the signals of phase transitions of the first order
associated with the divergence of capacities, it lacks the power to
reveal conveniently subtle points such as the emergence of the
second order phase transitions. It follows from (\ref{rngenthermo})
that we can treat the parameters $(r_0, q)$ as describing
independently the entropy and charge respectively. It is clear from
the temperature expression in (\ref{rngenthermo}) that in general
there are two threshold points associated with the divergence of the
specific heat at the fixed charge $Q$.  When the two minima of
$T(S)$ merge, the first-order phase transitions degenerate into a
second order one.  This occurs when $q=q'$ and $r_0=r_0'$, given by
\begin{equation}
q'^2=\fft{r_0'^{2(d-3)}}{(d-2)(2d-5)}\,,\qquad
r_0'^2=\fft{(d-3)^2}{(d-1)(d-2)\,\lambda^2}\,.
\end{equation}
When $q > q'$, the specific heat is always positive and
non-infinity, and hence there is no phase transition. When $q=q'$,
the specific heat is always positive and there is a second-order
phase transition occurs whenever
\begin{equation}
T=\fft{(d-3) \sqrt{(d-2)(d-1)}\lambda}{(2d-5)\pi}\,.
\end{equation}
For $0<q<q'$, there is a phase transition of the first order.

\section{Kerr-Newman-AdS black holes}

An analytical solution for the rotating charged black holes of
Einstein-Maxwell theories with or without cosmological constant is
only known for $d=4$, and it is called the Kerr-Newman-(AdS) black
hole \cite{carter}. Charged rotating black holes in five-dimensional
minimum gauged supergravity are also known \cite{cclp1}. However,
this theory contains an additional Chern-Simons term.

The thermodynamical properties for the Kerr-Newman-AdS black hole were
analyzed in \cite{cck}. The thermodynamical quantities are given by
\begin{eqnarray}
T&=&\fft{r_0^2(1 + (3r_0^2 + a^2)\lambda^2) - q^2-a^2}{4 \pi
r_0(r_0^2 + a^2)}\,,\qquad S=\fft{\pi (r_0^2 + a^2)}{\Xi}\,,\cr %%
\Omega&=&\fft{a(1 +\lambda^2 r_0^2)}{r_0^2 + a^2}\,,\quad
J=\fft{am}{\Xi^2}\,,\quad \Phi=\fft{q r_0}{r_0^2 + a^2}\,,\quad
Q=\fft{q}{\Xi}\,,\qquad M=\fft{m}{\Xi^2}\,,
\end{eqnarray}
where the parameters $m$ and $\Xi$ are given by
\begin{equation}
m=\fft{(r_0^2+a^2)(1 + \lambda^2 r_0^2) + q^2}{2r_0}\,,\qquad \Xi=1
- \lambda^2 a^2\,.
\end{equation}
The specific heat at constant angular momentum and charge is given
by \cite{cck}
\begin{equation}
C_{J,Q}=\fft{4\pi M T S}{1-4\pi T(2M + TS) + 2\lambda^2(Q^2 +
3\pi^{-1} S) + 6\pi^{-2}\lambda^4 S^2}\,.
\end{equation}
The Ruppeiner geometry of the Kerr-Newman black hole was studied in
\cite{rupp2}. The curvature is not sensitive to the Davies curve
where the specific heat $C_{J,Q}$ diverges.  It follows from the
previous discussion that this phase transition is expected to be
related to the curvature singularity of the free-energy metric.
Indeed, it is straightforward to verify that the denominator for the
Ricci scalar $R^{(F)}$ is the same as that of $C_{J,Q}$ up to
trivial non-vanishing factors.

In the following, we shall use $D(X)$ to denote the denominator of a
quantity $X$. The denominator of the Ricci scalar of the Weinhold
metric is given by
\begin{equation}
D(R^{(M)}) = 4m^2r_0^2 \zeta^2\,,
\end{equation}
where
\begin{equation}
\zeta =(r_0^2 + a^2)(1 + \lambda^2 r_0^2) (r_0^2 + a^2 +
\lambda^2r_0^2 (a^2 - 3r_0^2)) + q^2 (3a^2-r_0^2 - \lambda^2
r_0^2(r_0^2 + a^2))\,.
\end{equation}
 As one would have expected, this factor appears precisely in
the denominators of both charge capacitance $\widetilde
C_{T,\Omega}$ and moment of inertia $\widehat C_{T,\Phi}$, defined
by\footnote{We use $C,\widetilde C$ and $\widehat C$ to denote
specific heat, charge capacitance and moment of inertia,
respectively. The subscripts denote quantities that are held fixed.}
\begin{equation}
\widetilde C_{T,\Omega} = \fft{\partial Q}{\partial
\Phi}\Big|_{T,\Omega}\,,\qquad \widehat C_{T,\Phi} = \fft{\partial
J}{\partial \Omega}\Big|_{T,\Phi}\,.
\end{equation}
The denominator of the Ricci scalar in the Ruppeiner geometry is
\begin{equation}
D(R^{(S)}) = 4\pi^2 r_0(r_0^2 + a^2) T\,
D(R^{(M)})\,.\label{rwrelation}
\end{equation}

Note that there are a total of eight thermodynamical potentials
related by Legendre transformations. To systematically study
curvature singularities and capacity threshold points, we label them
as follows
\begin{eqnarray}
M\,, && \bar M = M - T S - \Phi Q - \Omega J\,,\cr %%
F=M - T S\,, && \bar F = M - \Phi Q - \Omega J\,,\cr %%
H=M - \Phi Q\,, && \bar H = M - T S - \Omega J\,,\cr %%
L=M - \Omega J\,,&& \bar L = M - T S - \Phi Q\,,
\end{eqnarray}
Following discussions in section 2, the barred and unbarred
quantities are conjugate pairs. There are a total of four
independent Ricci scalar quantities, and each one gives rise to one
singularity, resulting in total of four different singularities.
There are twelve capacities: four specific heat, four charge
capacitances, and four moments of inertia. There are eight
capacities that involve a total of four threshold points,
corresponding precisely the four curvature singularities. The
relationship between curvature singularities and capacity threshold
points can be summarized as follows
\begin{eqnarray}
R^{(\bar M)}=- R^{(M)} &\longrightarrow & \fft{\partial Q}{\partial
\Phi}\Big |_{T,\Omega}\sim \fft{\partial J}{\partial
\Omega}\Big|_{T,\Phi}\sim T \fft{\partial S}{\partial T}\Big
|_{\Phi,\Omega} \,,\cr %%
R^{(\bar F)}=-R^{(F)} &\longrightarrow & \fft{\partial M}{\partial
T}\Big |_{Q,J}\,,\cr %%
R^{(\bar H)}=-R^{(H)} &\longrightarrow & \fft{\partial J}{\partial
\Omega}\Big |_{T,Q}\sim T\fft{\partial S}{\partial T}\Big
|_{Q,\Omega}\,, \cr %%
R^{(\bar L)}=-R^{(L)} &\longrightarrow & \fft{\partial Q}{\partial
\Phi}\Big |_{T,J}\sim T\fft{\partial S}{\partial T}\Big|_{\Phi, J}
\,.%%
\end{eqnarray}
Here ``$\sim$'' links terms which have the same pole.  The
remaining quantities ${\partial Q}/{\partial \Phi}\Big|_{S, J}$,
${\partial Q}/{\partial \Phi}\Big|_{S, \Omega}$, ${\partial
J}/{\partial \Omega}\Big|_{S, \Phi}$ and ${\partial J}/{\partial
\Omega}\Big|_{S, Q}$ have no threshold points.  We see that all
possible phase transitions correspond to curvature singularities of
certain thermodynamical metrics. See appendix B for explicit
results.

\section{Kerr-AdS black holes}

General Kerr-AdS black holes were constructed in \cite{glpp}.
Their thermodynamical quantities were obtained in \cite{gpp}, which we
adopt for our calculation. Their general thermodynamical geometries are
complicated.  We shall consider only some special and simple cases.

\bigskip\bigskip
\noindent{\bf Four dimensions:}
\bigskip

Although this is a special case of the Kerr-Newman-AdS black hole
discussed in the previous section, it is instructive to list
it here.  The thermodynamical metrics are two-dimensional, and we
find the following scalar curvatures for the $F(T, J)$, $M(S, J)$
and $S(M,J)$ metrics:
\begin{equation}
R^{(F)} = \fft{P_1(r_0,a,\lambda)}{r_0\zeta_1^2}\,,\qquad R^{(M)}=
\fft{\lambda^2 P_2(r_0,a,\lambda)}{(r_0^2+a^2)\zeta_2^2}\,,
\qquad R^{(S)} = \fft{P_3(r_0,a,\lambda)}{4\pi^2(r_0^2+a^2) T\, \zeta_2^2}\,,
\end{equation}
where $P_i$ are non-singular polynomial functions of $r_0, a,
\lambda$, and
\begin{eqnarray}
\zeta_1 &=& 3a^4 + 6a^2 r_0^2 - r_0^4 + (a^6 + 13a^4 r_0^2 + 23 a^2
r_0^4 + 3r_0^6)\lambda^2 + a^2 r_0^2(a^2 + 3r_0^2)^2
\lambda^4\,,\cr%%
\zeta_2 &=& -r_0^2 -a^2 + \lambda^2 r_0^2
(3r_0^2-a^2)\,.\label{zeta12kerrads}
\end{eqnarray}
All specific heats and moments of inertia are given by
\begin{eqnarray}
C_J&\equiv& T\fft{\partial S}{\partial T}\Big|_J =\fft{8\pi r_0(r_0^2
+ a^2)^3 T}{(1-a^2\lambda^2)\zeta_1}\,,\qquad
C_\Omega\equiv T \fft{\partial S}{\partial T} \Big|_\Omega
=\fft{8\pi^2 r_0^3(r_0^2 + a^2) T}{(1-a^2\lambda^2) \zeta_2}\,,
\cr %%
\widehat C_T &\equiv& \fft{\partial J}{\partial \Omega}\Big|_T =
\fft{(r_0^2+a^2)\zeta_1}{2r_0(1-a^2\lambda^2)^3\zeta_2}\,,\qquad
\widehat C_S \equiv \fft{\partial J}{\partial \Omega}\Big |_S =
\fft{(r_0^2 + a^2)^3}{2r_0^3 (1-a^2\lambda^2)^3}\,,
\end{eqnarray}
The relationship between curvature singularities and
capacity thresholds is clear.

    Note that the $\lambda^2$ factor in $R^{(M)}$ implies that
the Ricci scalar for the Weinhold metric vanishes for zero
cosmological constant. It appears that the Weinhold metric fails to
predict the phase transition associated with the divergence of the
moment of inertia. However, as we can see from (\ref{zeta12kerrads})
that $\zeta_2$ has no zero when $\lambda=0$, and hence $\widehat
C_T$ will not diverge.

\bigskip\bigskip
\noindent{\bf Five dimensions:}
\bigskip

In this case, there are three conserved quantities, mass and two
angular momenta, parameterized by $r_0$ and $a$ and $b$.  The scalar
curvature for the Weinhold metric in this case diverges when
\begin{equation}
r_0^4(1 + \lambda^2 (a^2 + b^2-2r_0^2)) - a^2b^2=0\,.
\label{d5kerrwein}
\end{equation}
As in the previous case, this divergence coincides with the
divergence of the moments of inertia, defined by
\begin{equation}
\widehat C_{T,\Omega_b}\equiv \fft{\partial J_a}{\partial
\Omega_a}\Big|_{T,\Omega_b}\,,\qquad \widehat C_{T,\Omega_a}\equiv
\fft{\partial J_b}{\partial \Omega_b}\Big|_{T,\Omega_a}\,.
\end{equation}
They turn out to have the same divergent pole. For the Ruppeiner
geometry, we find the same relationship as (\ref{rwrelation}).

     A new feature arises in the free-energy metric.  There
are two singularities in the Ricci scalar $R^{(F)}$.  One is
associated with the divergence of the specific heat $C_{J_a, J_b}$.
The other is located at
\begin{equation}
a^2 b^2 + (a^2 + b^2) r^2 - 3 r_0^4 =0\,,
\end{equation}
which does not explicitly depend on the cosmological constant.  It
turns out that this point corresponds to the divergence of the
moments of inertia defined by
\begin{equation}
\widehat C_{S,\Omega_b}\equiv \fft{\partial J_a}{\partial
\Omega_a}\Big|_{S,\Omega_b}\,,\qquad \widehat C_{S,\Omega_a}\equiv
\fft{\partial J_b}{\partial \Omega_b}\Big|_{S,\Omega_a}\,.
\end{equation}
Both moments of inertia have the same divergent pole.

     To summarize, there are a total of eight thermodynamical potentials
related to each other by the Legendre transformations, given by
\begin{eqnarray}
M\,,&& \bar M = M - T S - \Omega_a J_a - \Omega_b J_b\,,\cr %%
F=M - TS\,,&& \bar F = M - \Omega_a J_a - \Omega_b J_b\,,\cr %%
L_a = M - \Omega_a J_a\,,&& \bar L_a = M - TS -
\Omega_b J_b\,,\cr %%
L_b = M - \Omega_b J_b\,, && \bar L_b = M - TS - \Omega_a J_a
\,.
\end{eqnarray}
The following are the relations between curvature
singularities and capacity thresholds:
\begin{eqnarray}
R^{(M)}=-R^{(\bar M)} &\longrightarrow & T\fft{\partial S}{\partial T}
\Big |_{\Omega_a,\Omega_b} \sim \fft{\partial J_a}{\partial \Omega_a}
\Big |_{T,\Omega_b} \sim \fft{\partial J_b}{\partial \Omega_b} \Big|
_{T,\Omega_a}\,,\cr %%
R^{(F)}=-R^{(\bar F)} &\longrightarrow & T\fft{\partial S}{\partial
T}\Big|_{J_a,J_b}\,,\qquad \fft{\partial J_a}{\partial \Omega_a}
\Big |_{S,\Omega_b}\sim \fft{\partial J_b}{\partial \Omega_b}
\Big|_{S, \Omega_a}\,,\cr %%
R^{(L_a)}=-R^{(-\bar L_a)} &\longrightarrow& \fft{\partial
J_a}{\partial \Omega_a}\Big |_{S, J_b}\,,\qquad T\fft{\partial
S}{\partial T}\Big|_{J_a, \Omega_b}\sim \fft{\partial J_b}{\partial
\Omega_b}\Big |_{T,J_a}\,,\cr %%
R^{(L_b)}=-R^{(-\bar L_b)} &\longrightarrow& \fft{\partial
J_b}{\partial \Omega_b}\Big |_{S, J_a}\,, \qquad T \fft{\partial
S}{\partial T}\Big|_{J_b, \Omega_a}\sim \fft{\partial J_a}{\partial
\Omega_a}\Big |_{T,J_b} \,.
\end{eqnarray}
The Weinhold metric and its conjugate pair have one
singularity whilst the remaining 6 metrics all have two
singularities. It is important to emphasize that all possible phase
transition signals, associated with the divergence of any of the
twelve capacities, are captured by curvature singularities.
Conversely, there is not a single curvature singularity that is not
related to a phase transition signal.  Detailed results can be
found in appendix C.

\bigskip\bigskip
\noindent{\bf Kerr-AdS with a single angular momentum:}
\bigskip

In this case, we take $a_1=a$ and $a_i=0$ for $i\ge 2$.  As in the
general $D=5$ case, there are two curvature singularities in the
metric associated with the free-energy.  One corresponds to the
divergence of the specific heat $C_J$, whilst the other occurs when
we have
\begin{eqnarray}
d=2N+1:&& r_0^2 = \fft{(2N-3)a^2}{2N-1}\,.\cr
d=2N+2:&& r_0^2 = \fft{(N-1) a^2}{N}\,.
\end{eqnarray}
This point corresponds to the divergence of the
moment of inertia
\begin{equation}
\widehat C_{S} = \fft{\partial J_a}{\partial \Omega_a}\Big |_S\,.
\end{equation}

The Weinhold and Ruppeiner metrics give no new results.  We have
$R^{(M)}\sim C_\Omega \sim \widehat C_T$ and $D(R^{(R)})=T D(R^{(M)})$.

\bigskip\bigskip
\noindent{\bf Kerr-AdS with all equal angular momenta:}
\bigskip

In this case, we take $a_i=a$ for all $i$. We find the same
correspondence between the free-energy metric and the specific
heat $C_J$.  There appears to be a correspondence between the
curvature singularity and the divergence of the capacitance $C_{S}$,
since both the Ricci scalar and the capacitance have the same factor
in the denominator, namely $(2N-1) r_0^2 + a^2$. However, this
factor can never be zero.  The result for the Weinhold and Ruppeiner
geometry is the same as that of previous examples.

\section{Black Ring}

The black ring solution in five dimensions with single angular
momentum was first obtained in \cite{emprea}. We shall adopt the notation of
\cite{lmp}, and the thermodynamical quantities are given by
\begin{eqnarray}
T &=& \fft{1}{2\pi} \eta_2 (\eta_1 + \eta_2)\,,\qquad
S=\fft{\pi^2}{2\eta_2 (\eta_1^2-\eta_2^2) (\eta_1 + \eta_2)^3}\,,\cr
\Omega &=& (\eta_1^2-\eta_2^2)\sqrt{\fft{\eta_2}{\eta_1}}\,, \qquad
J=\fft{\pi \eta_1^{\fft32}}{4\eta_2^{\fft32} (\eta_1^2 -
\eta_2)^2(\eta_1 + \eta_2)^2}\,,
\end{eqnarray}
where $\eta_1>\eta_2 >0$ and $\eta_1\eta_2<1$.  The Ruppeiner
geometry of this system was studied in \cite{blackringthermo}.

In this case, we find that the Weinhold metric is flat. This should
imply that there is no phase transition associated with the moment
of inertia at constant temperature. It is indeed the case, as we
shall see next. The Ricci scalars for the free-energy and Ruppeiner
metrics are given by
\begin{eqnarray}
R^{(F)} &=& - \fft{192 \eta_1\eta_2 (\eta_1-\eta_2)^2(\eta_1+\eta_2)^6
(\eta_1^2 - 2\eta_1 \eta_2 + 9 \eta_2^2)}{\pi (\eta_1^2 - 6\eta_1
\eta_2 - 3\eta_2^2)^2 (3\eta_1^2 + 2\eta_1\eta_2 + 3 \eta_2^2)^2}\,,\cr%%
R^{(S)} &=& - \fft{2\eta_2(\eta_1-\eta_2)^2(\eta_1 + \eta_2)^3
(\eta_1^2 - 6\eta_1\eta_2 - 3 \eta_2^2)}{\pi^2
(\eta_1-3\eta_2)^2}\,.
\end{eqnarray}
Each gives a singularity, corresponding to $\eta_1^2 - 6\eta_1\eta_2
- 3\eta_2^2=0$ or $\eta_1=3\eta_2$.  These points should be related
to certain phase transitions.  For the Emparan-Reall black ring, one
can readily calculate both the specific heat at constant angular
momentum $J$ and angular velocity $\Omega$,
\begin{eqnarray}
C_J &=& \fft{3\pi^2 (\eta_1 - 3\eta_2)}{2(\eta_1 -
\eta_2)\eta_2(\eta_1 + \eta_2)^3 (3\eta_1^2 + 2 \eta_1 \eta_2 + 3
\eta_2^2)}\,,\cr%%
C_\Omega &=& \fft{\pi^2 (\eta_1^2 - 6\eta_1\eta_2 - 3 \eta_2^2)}{2
\eta_2 (\eta_1 - \eta_2) (\eta_1 + \eta_2)^6}\,,
\end{eqnarray}
and moments of inertia at constant $T$ or $S$:
\begin{eqnarray}
\widehat C_T  &=& - \fft{\pi\eta_1^2 (3\eta_1^2 + 2 \eta_1\eta_2 +
3\eta_2^2)}{4\eta_2^2 (\eta_1-\eta_2)^3 (\eta_1 + \eta_2)^7} \,,\cr%%
\widehat C_S &=&  - \fft{3\pi\eta_1^2 (\eta_1-3\eta_2)}{4(\eta_1 -
\eta_2)^3 \eta_2^2 (\eta_1 + \eta_2)^4 (\eta_1^2 - 6\eta_1 \eta_2 -
3\eta_2^2)}\,.
\end{eqnarray}
We note that $\widehat C_T$ has neither pole nor zero,
consistent with the flat Weinhold geometry. The singularity of
$R^{(F)}$ corresponds to the divergence of $\widehat C_S$, not
$C_J$.  This is very different from previous black hole
thermodynamics.  Although $R^{(F)}$ and $C_J$ share a common
denominator, it does not vanish.  Another new feature is that
curvature singularity of the Ruppeiner metric corresponds to the
zero of $C_J$ and $\widehat C_S$, instead of a typical divergent
point.

\section{Conclusions}

A geometrical understanding of the state space of a thermodynamical
system has been proposed for thirty years.  Its application in
black hole thermodynamics has generated interesting results.  An
important and difficult question is how to construct an appropriate
metric over the space. People have been focusing on mainly two
metrics, the energy metric proposed by Weinhold and the entropy
metric proposed by Ruppeiner.

By investigating various black holes, we find that although the
energy and entropy metrics can reveal certain phase transitions at
their curvature singularities, they miss some important ones,
including the phase transition along the Davies curve where the
specific heat with constant charge and angular momentum diverges. We
generalize this procedure and consider other thermodynamical
potentials via Legendre transformations as the seed function for the
Hessian matrix. We find that the curvature singularity of the
free-energy metric is always located along the Davies curve.  We
obtain the full set of curvature singularities for all the metrics
generated by thermodynamical potentials.  We find that they form
conjugate pairs: the metrics are negative of each other.

A black hole in a supergravity theory can typically involve many
conserved quantities, including the mass, charges and angular
momenta. There are three types of capacities: the specific heat,
charge capacitances and momenta of inertia.  There may exist a few
threshold points where these capacities change sign, signaling a
change of stability and phase transitions.  For all black holes and
a black ring we have examined, we find that the full collection of
these thresholds is exactly the same as the full collection of the
curvature singularities. We expect that this is true for all
thermodynamical systems.  On the other hand, we did not find a
thermodynamical metric which could signal all of the thermodynamic
phase transitions. This should not be surprising since different
thermodynamical phase transitions occur in very different ensembles.
It is reasonable to expect that only the metric of an appropriate
thermodynamical potential is relevant to a particular phase
transition.

A thermodynamical system with $n$ pairs of intensive/extensive
variables $(\mu_i, N_i)$ have an $n$-dimensional state space that
can be viewed as an embedding space of $2n$-dimensional flat space
with $(\mu_i,N_i)$ as coordinates. There exist a total of $2^{n-1}$
inequivalent thermodynamical potential metrics based on the Hessian
matrix, which are in turn possible to embed in flat
$(n,n)$-dimensional space. This provides a rather simple geometric
picture of thermodynamical state space. Furthermore it also provides
a very simple mechanism to compute the metrics and curvature
invariants. The questions of the physical significance of these
metrics and what is the principle underling their relationship with
phase transitions remain open.

\section*{Acknowledgement}

We are grateful to Jim Ferguson and Pang Yi for useful discussions,
and to KITPC for hospitality. We are also grateful to the referee
for proofreading our manuscript. H.~Liu and M.~Luo are supported in
part by the National Science Foundation of China (10425525,
10875103), National Basic Research Program of China (2010CB833000),
and Zhejiang Uni-versity Group Funding (2009QNA3015).

\appendix
\section{The Van der Waals Model}

In this appendix, we consider the thermodynamical geometry for the
Van der Waals model. Its similarity with the black hole
thermodynamics was noted in \cite{Sahay:2010wi}.  We shall
demonstrate, as in the case of black holes, additional
thermodynamical metrics are necessary to probe all the phase
transitions.  Adopting the notation in \cite{Sahay:2010wi}, the
free-energy for the Van der Waals model is given by
\begin{equation}
F=N (-c_v T\log T - \zeta T + \epsilon) - N T \log (e (V - N b) N^{-1}) -
N^2 a V^{-1}\,,
\end{equation}
where $F$ is a function of $T$ and $V$, satisfying
\begin{equation}
dF = - S dT + P dV\,.
\end{equation}
For simplicity, we set $\zeta=1$ and $\epsilon=1$.  The entropy and
pressure are given by
\begin{equation}
S=N (c_v (1 + \log T) + \log (e(V-N b) N^{-1}))\,,\qquad
P=N\Big(\fft{T}{N b - V} + \fft{N a}{V^2}\Big)\,.
\end{equation}
The $(P,V)$ pair is very much like the $(\Phi, Q)$ or $(\Omega, J)$
pairs in black hole physics.  There are a total of four types of
capacities, given by
\begin{eqnarray}
C_V&=& T\fft{\partial S}{\partial T}\Big|_V=c_v\,,\cr %%
C_P &=& T\fft{\partial S}{\partial T}\Big|_P=
\fft{N ((1+c_v) TV^3- 2 a c_v N (Nb - V)^2)}{TV^3 - 2Na (Nb - V)^2}\,,\cr %%
\fft{\partial V}{\partial P}\Big |_T &=& \fft{(N b - V)^2 V^3}{Na (TV^3 - 2Na
(Nb - V)^2)}\,,\cr %%
\fft{\partial V}{\partial P}\Big |_S &=& \fft{c_v V^3 (N b -
V)^2}{Na( (1+c_v) TV^3 - 2a c_v N (Nb - V)^2)}\,.
\end{eqnarray}
Similar to RN AdS black holes, the system has two threshold points.
One difference is that the $C_V$ here is a constant, whilst the
corresponding $C_Q$ of an RN black hole has a divergent point,
associated with the Davies curve.  Note that since the factor
$(Nb-V)^2$ is non-negative, the vanishing of this factor does not
correspond to a threshold point. Also we find that both Ruppeiner
and Weinhold curvature reveals only one threshold, associated with
$C_P$, as in the case of the black hole cases discussed in the
paper. To be specific, we have
\begin{eqnarray}
R^{(M)} &=&  \fft{aV^3 (N b - V)^2}{c_v (TV^3 - 2 N a (N b - V)^2)^2}\,,\cr%%
R^{(S)} &=& - \fft{2a (N b - V)^2 (TV^3 - Na (Nb - V)^2)}{c_v
(TV^3 - 2 N a (Nb - V)^2)^3}\,.
\end{eqnarray}
The other divergent point can be revealed by the free-energy metric,
{\it i.e.}
\begin{equation}
R^{(F)} = \fft{a c_v (3N b - 2V) V^2 (N b - V)^2}{
\Big((1 + c_v) T V^3 - 2a c_v N (N b - V)^2\Big)^2}\,.
\end{equation}

\section{Capacities of Kerr-Newman AdS Black holes}

In this appendix, we present some detailed results that were discussed
in section 4.  The thermodynamical curvatures are in general too
complex to present.  Here we only give the poles:
\begin{eqnarray}
R^{(M)}=-R^{(\bar M)} \sim \fft{1}{\zeta_1^2}\,,&&
R^{(F)}=-R^{(\bar F)} \sim \fft{1}{\zeta_2^2}\,,\cr %%
R^{(L)}=-R^{(\bar L)} \sim \fft{1}{\zeta_3^2}\,,&&
R^{(H)}=-R^{(\bar H)} \sim \fft{1}{\zeta_4^2}\,,
\end{eqnarray}
where
\begin{eqnarray}
\zeta_1 &=& -(r_0^2 + a^2)(1 + \lambda^2 r_0^2)
(r_0^2 + a^2 + \lambda^2 r_0^2 (a^2 - 3r_0^2)) + q^2(r_0^2-3a^2 +
\lambda^2 r_0^2 (r_0^2 + a^2))\,,\cr %%
\zeta_2 &=&  3 a^4+q^4 \left(3+a^2 \lambda ^2\right)+16 a^4 \lambda ^2
r_0^2+19 a^4 \lambda ^4 r_0^4+6 a^4 \lambda ^6 r_0^6+a^6
\left(\lambda +\lambda ^3 r_0^2\right){}^2\cr %%
&&+r_0^4 \left(-1+2 \lambda ^2 r_0^2+3 \lambda ^4 r_0^4\right)+a^2 r_0^2
\left(6+29 \lambda ^2 r_0^2+32 \lambda ^4 r_0^4+9 \lambda ^6 r_0^6\right)\cr %%
&&+ q^2 \left(a^4 \left(2 \lambda ^2-2 \lambda ^4 r_0^2\right)+
2 \left(r_0^2+3 \lambda ^2 r_0^4\right)+a^2 \left(6+8 \lambda ^2 r_0^2-
6 \lambda ^4 r_0^4\right)\right)\,, \cr %%
\zeta_3 &=& a^8 \lambda^2\left(1 +\lambda ^2 r_0^2\right){}^2+
a^6 \left(3+4 q^2 \lambda ^2+17 \lambda ^2 r_0^2+21 \lambda ^4 r_0^4
+7 \lambda ^6 r_0^6\right)\cr %%
&&+r_0^2 \left(q^4-r_0^4+4 q^2 \lambda ^2 r_0^4+2 \lambda ^2 r_0^6
+3 \lambda ^4 r_0^8\right)\cr &&+3 a^4 \left(q^4 \lambda ^2+4 q^2
\left(1+3 \lambda ^2 r_0^2+\lambda ^4 r_0^4\right) +r_0^2 \left(3+15
\lambda ^2 r_0^2+17 \lambda ^4 r_0^4+5 \lambda ^6
r_0^6\right)\right) \cr %%
&&+a^2 \left(3 q^4 \left(3+\lambda ^2 r_0^2\right)+12 q^2
\left(r_0^2+3 \lambda ^2 r_0^4+\lambda ^4 r_0^6\right)\right.\cr
&&\left.+r_0^4 \left(5+31 \lambda ^2 r_0^2+ 35 \lambda ^4 r_0^4+9
\lambda ^6 r_0^6\right)\right)\,,\cr%%
\zeta_4&=& -a^4 \left(1+\lambda ^2 r_0^2\right){}^2+r_0^2
\left(1+\lambda ^2 r_0^2\right) \left(3 q^2-r_0^2+
3 \lambda ^2 r_0^4\right)\cr %%
&&-a^2 \left(2 r_0^2-2 \lambda ^4 r_0^6+q^2 \left(1-3 \lambda ^2
r_0^2\right)\right)\,.
\end{eqnarray}
There are a total of twelve capacities: four specific heat, four
charge capacitances and four moments of inertia.  It turns out that
the vanishing points of these $\zeta_i$'s are precisely the threshold
points of capacities. Each specific heat involves one $\zeta_i$.
Explicitly,
\begin{eqnarray}
C_{Q,J} &=& T \fft{\partial S}{\partial T}\Big |_{Q,J} =
\frac{8 \pi ^2 r_0 \left(a^2+r_0^2\right){}^2
\left(q^2+r_0^2+\lambda ^2 r_0^4+a^2 \left(1+\lambda ^2 r_0^2\right)
\right)T}{(1-a^2 \lambda ^2)\,\zeta_2}\,\cr %%
C_{Q,\Omega}&=& T \fft{\partial S}{\partial T}\Big |_{Q,\Omega} =
\frac{8 \pi ^2 r_0^3 \left(a^2+r_0^2\right){}^2
\left(1+\lambda ^2 r_0^2\right)T}{\left(1-a^2
\lambda ^2\right)\zeta_4}\,,\cr %%
C_{\Phi,J} &=& T \fft{\partial S}{\partial T}\Big |_{\Phi,J}=
\frac{8\pi^2 r_0 (r_0^2 + a^2)^2 T}{\left(1-a^2 \lambda
^2\right)\zeta_3}\Big(a^4 \left(1+\lambda ^2 r_0^2\right)\cr
&&\qquad+r_0^2 \left(q^2+r_0^2+\lambda ^2 r_0^4\right)+a^2 \left(q^2
\left(3+2 \lambda ^2 r_0^2\right) +2 \left(r_0^2+\lambda ^2
r_0^4\right)\right)\Big)\,,\cr %%
C_{\Phi, \Omega} &=& T \fft{\partial S}{\partial T}\Big |_{\Phi,\Omega}
= \frac{8 \pi ^2 r_0^3 \left(a^2+r_0^2\right){}^2
\left(1+\lambda ^2 r_0^2\right)T}{\left(1-a^2 \lambda^2\right)\zeta_1}
\end{eqnarray}
Only two charge capacitances and two moments of inertia involve $\zeta_i$,
\begin{eqnarray}
\widetilde C_{T,J} &=& \fft{\partial Q}{\partial \Phi}\Big |_{T,J}=
\frac{\left(a^2+r_0^2\right){}^2\zeta_2}{ \left(1-a^2\lambda
^2\right) r_0\,\zeta_3}\,,\cr \widetilde C_{T,\Omega}&=&
\fft{\partial Q}{\partial \Phi}\Big
|_{T,\Omega}=\frac{\left(a^2+r_0^2\right)\zeta_4}{\left(1-a^2
\lambda ^2\right) r_0\zeta_1}\,,\cr%%
\widehat C_{T,Q} &=& \fft{\partial J}{\partial \Omega}\Big |_{T,Q}=-
\frac{\left(a^2+r_0^2\right){}^2\zeta_2}{2
\left(1-a^2 \lambda ^2\right)^3 r_0\zeta_4}\,,\cr %%
\widehat C_{T,\Phi} &=& \fft{\partial J}{\partial \Omega}\Big |_{T,\Phi}
=-\frac{\left(a^2+r_0^2\right)\zeta_3}{2 \left(1-a^2
\lambda ^2\right)^3 r_0\zeta_1}\,.
\end{eqnarray}
Note that $\zeta_2$ and $\zeta_3$ can appear in numerators as
well. The charge capacitances $\widetilde C_{S, J}$, $\widetilde
C_{S,\Omega}$ and moments of inertia $\widehat C_{S,Q}$, $\widehat
C_{S,\Phi}$ do not involve $\zeta_i$.  They have neither zero nor
pole and we shall not present them.

\section{Capacities of Kerr-AdS$_5$ Black holes}

In this appendix, we present some detailed results discussed
in section 5.  The thermodynamical curvatures are in general too
complex to present.  Here we only give the poles:
\begin{eqnarray}
R^{(M)}=-R^{(\bar M)} \sim \fft{1}{\zeta_1^2}\,,&& R^{(F)}=-R^{(\bar
F)} \sim \fft{1}{\zeta_2^2 \zeta_3^2}\,,\cr R^{(L_a)}=-R^{(\bar
L_a)}\sim\fft{1}{\zeta_4^2 \zeta_6^2}\,,&& R^{(L_b)}=-R^{(\bar
L_b)}\sim \fft{1}{\zeta_5^2 \zeta_7^2}\,,
\end{eqnarray}
where
\begin{eqnarray}
\zeta_1&=& a^2 b^2-\left(1+a^2 \lambda ^2+b^2 \lambda ^2\right)
r_0^4+2 \lambda ^2 r_0^6\,,\cr %%
\zeta_2&=& -a^2 b^2-\left(a^2+b^2\right) r_0^2 +3 r_0^4\,,\cr %%
\zeta _3 &=& 5 a^4 b^4+9 a^4 b^2 r_0^2+9 a^2 b^4 r_0^2+ 20 a^2 b^2
r_0^4+3 a^2 r_0^6+3 b^2 r_0^6-r_0^8\cr &&+a^2 b^2 \lambda ^6 r_0^4
\left(a^2+r_0^2\right) \left(b^2+r_0^2\right) \left(3
\left(a^2+b^2\right)+10 r_0^2\right)\cr &&+\lambda ^2 \left(3 a^4
b^4 \left(a^2+b^2\right)+3 a^2 b^2 \left(a^4+10 a^2 b^2+b^4\right)
r_0^2 \right.\cr &&\left.+39 a^2 b^2 \left(a^2+b^2\right) r_0^4+2
\left(2 a^4+33 a^2 b^2+2 b^4\right) r_0^6\right.\cr &&\left.+14
\left(a^2+b^2\right) r_0^8+2 r_0^{10}\right) +\lambda ^4 \left(a^6
b^6+9 a^4 b^4 \left(a^2+b^2\right) r_0^2\right.\cr &&\left.+a^2 b^2
\left(9 a^4+46 a^2 b^2+9 b^4\right) r_0^4+\left(a^2+b^2\right)
\left(a^4+43 a^2 b^2+b^4\right) r_0^6\right.\cr &&\left.+\left(3
a^4+49 a^2 b^2+3 b^4\right) r_0^8+ 6 \left(a^2+b^2\right)
r_0^{10}\right)\,,\cr %%
\zeta _4 &=& -a^2 \left(1-b^2 \lambda ^2\right)+
\left(3+b^2 \lambda ^2\right) r_0^2\,,\cr %%
\zeta _5&=&-b^2 \left(1-a^2 \lambda ^2\right)+
\left(3+a^2 \lambda ^2\right) r_0^2\,,\cr %%
\zeta _6&=& -b^2 \left(a^2+r_0^2\right) \left(3 a^2+3 a^2 \lambda ^4 r_0^4+
\lambda ^2 \left(a^4+8 a^2 r_0^2+r_0^4\right)\right)\cr %%
&&+r_0^4 \left(3 a^2-r_0^2+2 \lambda ^2 \left(2 a^4+7 a^2 r_0^2+
r_0^4\right)+a^2 \lambda ^4 \left(a^4+3 a^2 r_0^2+6 r_0^4
\right)\right)\,,\cr %%
\zeta _7 &=& -a^2 \left(b^2+r_0^2\right) \left(3 b^2+3 b^2 \lambda
^4 r_0^4+\lambda ^2 \left(b^4+8 b^2 r_0^2+r_0^4\right)\right)\cr
&&+r_0^4 \left(3 b^2-r_0^2+2 \lambda ^2 \left(2 b^4+7 b^2
r_0^2+r_0^4\right)+b^2 \lambda ^4 \left(b^4+3 b^2 r_0^2+6
r_0^4\right)\right)\,.
\end{eqnarray}
Note that $\zeta_1$, $\zeta_2$ and $\zeta_3$ are symmetric under the
interchange of $a$ and $b$, whilst $\zeta_4$ and $\zeta_5$,
$\zeta_6$ and $\zeta_7$ interchange with each other.

There are four specific heat, given by
\begin{eqnarray}
C_{J_a,J_b} &=& \frac{\pi ^3 \left(3+b^2 \lambda ^2+a^2 \lambda ^2 +
\left(1-b^2 \lambda ^2\right)\right) \left(a^2+r_0^2\right){}^3
\left(b^2+r_0^2\right){}^3T}{\left(1-a^2 \lambda ^2\right)
\left(1-b^2 \lambda ^2\right)\zeta _3}\,,\cr %%
C_{J_a,\Omega_b} &=&  \frac{T\zeta_5\pi ^3
\left(a^2+r_0^2\right){}^3 \left(b^2+r_0^2\right)}{\zeta _6
\left(1-a^2 \lambda ^2\right) \left(1-b^2 \lambda ^2\right)}
\,,\cr %%
C_{\Omega_a,J_b} &=&\frac{T\zeta_4\pi ^3 \left(a^2+r_0^2\right)
\left(b^2+r_0^2\right){}^3}{\zeta _7\left(1-a^2 \lambda ^2\right)
\left(1-b^2 \lambda ^2\right)}\,,\cr %%
C_{\Omega_a,\Omega_b} &=& \frac{T \zeta _2\pi ^3 \left(a^2+r_0^2\right)
\left(b^2+r_0^2\right)}{\zeta _1\left(1-a^2 \lambda ^2\right)
\left(1-b^2 \lambda ^2\right)}\,.
\end{eqnarray}
There are four moments of inertia associated with index $a$. They are given by
\begin{eqnarray}
\widehat C_{T,J_b} &=& \frac{\zeta _3\pi  \left(a^2+r_0^2\right)
\left(b^2+r_0^2\right)}{4 \zeta _7\left(1-a^2 \lambda ^2\right)^3
\left(1-b^2 \lambda ^2\right) r_0^2}\,,\cr %%
\widehat C_{T,\Omega_b} &=& \frac{\zeta _6 \pi  \left(a^2+r_0^2\right)
\left(b^2+r_0^2\right)}{4 \zeta _1 \left(1-a^2 \lambda ^2\right)^3
\left(1-b^2 \lambda ^2\right) r_0^2}\,,\cr %%
\widehat C_{S, J_b} &=&  \frac{\pi  \left(3+b^2 \lambda ^2+a^2
\lambda ^2
\left(1-b^2 \lambda ^2\right)\right) \left(a^2+r_0^2\right){}^3
\left(b^2+r_0^2\right)}{4 \zeta _4 \left(1-a^2 \lambda ^2\right)^3
\left(1-b^2 \lambda ^2\right) r_0^2}\,,\cr %%
\widehat C_{S, \Omega_b} &=& \frac{\zeta _5 \pi
\left(a^2+r_0^2\right){}^3 \left(b^2+r_0^2\right)}{4\zeta _2
\left(1-a^2 \lambda ^2\right)^3 \left(1-b^2 \lambda ^2\right)
r_0^2}\,.
\end{eqnarray}
The moments of inertia associated with index $b$ are given above,
but with $(a,b)$, $(\zeta_4,\zeta_5)$ and $(\zeta_6,\zeta_7)$
interchanged.

\end{document}